\documentclass[%
 reprint,
 amsmath,amssymb,
 aps,
floatfix,
]{revtex4-2}

\usepackage{amsmath,amsfonts,amssymb,bm,cancel,multirow,dsfont}
\usepackage{graphicx,float} 
\usepackage{enumitem}
\usepackage{color}
\usepackage{latexsym}
\usepackage{wasysym}
\usepackage{soul}
\usepackage{dcolumn}%
\usepackage{xcolor}
\usepackage{placeins}
\usepackage{ bbold }
\usepackage[]{changes}

\usepackage{layouts}%
\usepackage{comment}

\newcommand{\xm}[0]{{x_{\text{min}}}}
\newcommand{\xmavg}[0]{{\langle x_{\text{min}}\rangle}}
\newcommand{\xaavg}[0]{\langle x_{a}\rangle }
\newcommand{\argmax}[0]{\text{argmax} }
\newcommand{\dup}[0]{\mathop{}\!\mathrm{d}}
\newcommand{\pp}[2]{\frac{\partial #1}{\partial#2}}
\newcommand{\dd}[2]{\frac{\dup #1}{\dup #2}}
\newcommand{\lambW}[0]{\mathcal{W}}

\newcommand{\betaEPM}[0]{1.5}

\newcommand{\tens}[1]{ \underline{\underline{#1}}}
\newcommand{\tr}{\text{tr}}
\newcommand{\vc}[1]{\underline{#1}}

\newcommand{\tildet}[0]{\tilde{t}}
\newcommand{\tD}[0]{\tilde{D}}
\newcommand{\tgdot}[0]{\tilde{\dot{\gamma}}}
\newcommand{\tGamma}[0]{\tilde{\Gamma}}
\newcommand{\tnu}[0]{\tilde{\nu}}

\begin{document}

\title{Thermally activated intermittent flow in amorphous solids}
\author{Daniel Korchinski}
\email{djkorchi@phas.ubc.ca}
\affiliation{
 Department of Physics and Astronomy and Stewart Blusson Quantum Matter Institute,  University of British Columbia, Vancouver BC V6T 1Z1, Canada}%
\author{J\"org Rottler}%
\affiliation{
 Department of Physics and Astronomy and Stewart Blusson Quantum Matter Institute,  University of British Columbia, Vancouver BC V6T 1Z1, Canada}%

\begin{abstract}
Using mean field theory and a mesoscale elastoplastic model, we analyze the steady state shear rheology of thermally activated amorphous solids. At sufficiently high temperature and driving rates, flow is continuous and described by well-established rheological flow laws such as Herschel-Bulkley and logarithmic rate dependence. However, we find that these flow laws change in the regime of intermittent flow, were collective events no longer overlap and serrated flow becomes pronounced.  In this regime, we identify a thermal activation stress scale, $x_{a}(T,\dot{\gamma})$, that wholly captures the effect of driving rate $\dot{\gamma}$ and temperature $T$ on average flow stress, stress drop (avalanche) size and correlation lengths. Different rheological regimes are summarized in a dynamic phase diagram for the amorphous yielding transition. Theoretical predictions call for a need to re-examine the rheology of very slowly sheared amorphous matter much below the glass transition.

\end{abstract}

\maketitle

\section{Introduction}
\begin{figure*}
    \centering
    \includegraphics[width=\linewidth]{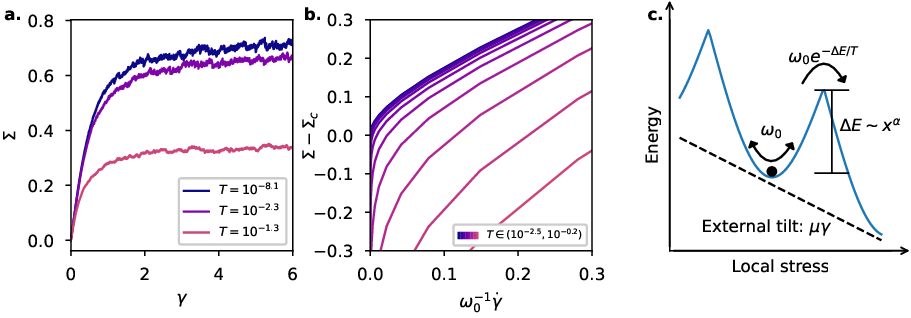}
    \caption{\textbf{a.} Stress-strain relation of ductile failure in amorphous solids for different temperatures (EPM simulation). \textbf{b.} Flow curves for different driving rates at different temperatures (EPM simulations). \textbf{c.} Schematic of energy landscape for a single soft-spot and Arrhenius activation at finite-temperature.  }
    \label{fig:schematic-failure}
\end{figure*}

Amorphous solids are a diverse class of materials, unified by a lack of long-range structural order in their constituent particles. 
Despite their widely disparate length scales (e.g.~metallic glasses $\approx 1$\AA\, versus foam rafts $\approx 1$cm), these materials have similar yielding and failure behavior~\cite{barrat_modeling_2007,nicolas_deformation_2018,berthier_yielding_2024}. After an initially Hookean elastic response to strain, these systems tend to either fail in a brittle manner, developing system-spanning shear bands and an abrupt collapse, or in a ductile manner, with a relatively smooth transition to a flowing state (cf. fig.~\ref{fig:schematic-failure}a)~\cite{divoux_ductile--brittle_2023}. This commonality arises because the response to deformation is enabled by swift, localized shear transformations (ST) that 
propagate stress across the system, stabilizing some regions and destabilizing others. Destabilized regions can in turn have their own STs, leading to ``avalanches'' of collective rearrangements. In the limit of athermal, and slow driving,  these collective effects become most pronounced, leading to a form of self-organized criticality \cite{talamali_strain_2012,lin_scaling_2014,lin_density_2014,lin_criticality_2015,lin_microscopic_2018,korchinski_signatures_2021,jocteur_yielding_2024,tyukodi_avalanches_2019,tyukodi_diffusion_2018,jagla_avalanche-size_2015,fernandez_aguirre_critical_2018,ferrero_properties_2021,eferrero_criticality_2019,budrikis_universal_2017},. These avalanches organize to be scale-free and can be described with a non-trivial set of exponents. The stress fluctuates strongly about some critical stress, $\Sigma_c$, which manifests as macroscopic serrations in the stress-strain curve (cf. fig.~\ref{fig:schematic-failure}a). 

The picture changes somewhat if the system is driven more quickly. At higher shear rate $\dot{\gamma}$, the loading becomes so rapid that new avalanches are initiated  before old avalanches have finished dissipating energy, meaning that the overall stress increases above the critical flow stress $\Sigma_c$ (cf. fig.~\ref{fig:schematic-failure}b). This increase in flow-stress is typically described by the Herschel-Bulkley (HB) law for yield stress fluids,
\begin{equation}
    \Sigma = \Sigma_c + K \dot{\gamma}^n \label{eq:hb_lowest_order}
\end{equation}
with $n$ the Herschel-Bulkley exponent (typically $n<1$, i.e shear-thinning behavior) and $K$ some material dependent constant. In this regime, which we term the {\em continuous flow regime}, stress fluctuations are reduced as the now spatiotemporally overlapping avalanches become increasingly decorrelated~\cite{clemmer_criticality_2021,clemmer_criticality_2021-1}. In analogy with second-order phase-transitions, the correlation length decreases $\xi \sim (\Sigma-\Sigma_c)^{-\nu}$, and in a similar manner, the approach to criticality can be written as a recast HB relation, $\dot{\gamma} \sim (\Sigma-\Sigma_c)^\beta$, with an order parameter exponent $\beta = 1/n$~\cite{liu_driving_2016,karimi_inertia_2017}. 

While thermal fluctuations are largely irrelevant for driven amorphous solids with large constituent particles, they become important for molecular and metallic glasses that are operated closer to their glass transition. In the case of externally driven solids, temperature has the effect of reducing the flow stress $\Sigma$ and also the size of fluctuations (cf. fig.~\ref{fig:schematic-failure}b)~\cite{khanouki_temperature_2021}. Additionally, temperature allows for the spontaneous relaxation of otherwise mechanically stable configurations of the amorphous solid, meaning that systems held at fixed $\Sigma<\Sigma_c$ will now  exhibit extremely slow flow referred to as ``creep'', with a complex history dependent transient, before reaching a very slow steady-state flow~\cite{liu_creep_2018,liu_elastoplastic_2021,xu_yielding_2021,popovic_scaling_2021,popovic2021thermalflow}. 

It was suggested in refs.~\cite{popovic2021thermalflow,ferrero_yielding_2021} that the crossover between athermal (i.e.~deformation dominated activation) and thermal (i.e.~spontaneous activation dominates) flow can be described with a Widom scaling ansatz, so $\Sigma-\Sigma_c \sim T^{1/\alpha}f(\dot{\gamma}/T^{\psi})$ with $\psi = \beta/\alpha$, implying $\dot{\gamma}_{cross} \sim T^{\beta/\alpha}$. This relation can be obtained from a simple scaling argument: in athermal systems external driving introduces a stress-scale $\Delta\Sigma=\Sigma-\Sigma_c\sim \dot{\gamma}^{1/\beta}$, while temperature introduces a natural stress scale $x_a\sim T^{1/\alpha}$. Naturally one expects a rheological crossover occurs when these two scales compete, i.e. when $T^{1/\alpha} \sim \dot{\gamma}^{1/\beta}$. Both a mean-field Fokker-Planck type equation analysis \cite{popovic2021thermalflow} and simulations of an elastoplastic model (EPM) \cite{popovic2021thermalflow,ferrero_yielding_2021} support this argument.  However, other work also using EPM found a satisfactory numerical collapse for a crossover scaling as $\dot{\gamma}_{cross} \sim ^{1/\alpha}$ \cite{korchinski_dynamic_2022}. A first goal of this paper is to clarify this point of disagreement and thus the rheological flow laws in the continuous regime.

As we will show, resolution of this question requires appreciation of an additional flow regime with different physics. For not too high driving rates and temperatures, it is possible to enter a regime where avalanches are still separated in space and time, but can be nucleated via a competition of mechanical and thermal driving.  In this thermal {\em intermittent regime}, both scaling properties of the avalanches and the flow curves are distinct from athermal quasistatic deformation and continuous flow. A second goal of this paper is to develop a framework for understanding the statistical properties of collective events and rheology in this underexplored intermittent regime. A central component will be the concept of a thermal activation scale $x_a(T,\dot{\gamma})$, which is the typical threshold for a stable site to yield via thermal activation given deformation at fixed temperature and shear rate.

In what follows, we analyze the steady state shear rheology of thermally activated amorphous solids and the scaling behavior of avalanches with two complementary modelling approaches executed side by side: a mesoscale two-dimensional elastoplastic model, as well as a mean-field Fokker-Planck description that is a straightforward extension of the Hebraud-Lequeux (HL) model \cite{hebraud_mode-coupling_1998}. We show that mean-field theory captures well the crossover between athermal and thermal flow in the continuous regime, and, perhaps surprisingly, can be adapted to describe the rheology in the intermittent regime as well. We develop analytical expressions for the activation scale that rationalizes avalanche size distributions, correlation lengths, and rheology in the intermittent regime. Our development culminates in a comprehensive dynamical phase diagram in the temperature - shear rate plane, with phase boundaries that delineate crossovers between the different rheological regimes.

\section{Numerical modelling}
The treatment of amorphous yielding has been approached at different levels of coarse-graining. While large-scale direct molecular dynamics simulations are possible, when probing the universal behaviour of a system near its phase transition, it is sufficient to use a meoscopic model with the right symmetries. To that end, the essential features captured in elastoplastic models, are that amorphous solids plastically yield by through localized STs and that these rearrangements trigger a long-ranged stress redistribution in the system, as predicted by elasticity theory. To that end, EPMs generally consist of a lattice of sites, with disordered local yield thresholds~\cite{baret_extremal_2002,homer_mesoscale_2009}. While EPMs models have been widely used to universal physics in brittle yielding~\cite{rossi_finite-disorder_2022,fielding_yielding_2021}, and ductile yielding~\cite{lin_scaling_2014,lin_density_2014,lin_criticality_2015,korchinski_signatures_2021,jocteur_yielding_2024,tyukodi_avalanches_2019,tyukodi_diffusion_2018,jagla_avalanche-size_2015,fernandez_aguirre_critical_2018,ferrero_properties_2021,eferrero_criticality_2019,budrikis_universal_2017}, they have also been used to describe non-universal physics, like memory effects~\cite{kumar_mapping_2022,khirallah_yielding_2021}, vibrated yielding~\cite{le_goff_criticality_2019,goff_giant_2020}, driven shear band formation~\cite{martens_spontaneous_2012}, and can be closely matched to atomistic simulations~\cite{liu_elastoplastic_2021}. For a review of EPMs and the yielding transition, see~\cite{nicolas_deformation_2018}.  

Our implementation of the EPM (aee also ref.~\cite{korchinski_dynamic_2022}) consists of a two-dimensional square grid of $L^2$ lattice sites. Each site has a yield threshold $\sigma_{th,i}$ drawn from a Weibull distribution of unit mean and shape-parameter $k=2$ as suggested by molecular dynamics~\cite{barbot_local_2018,castellanos_history_2022}. When a site is mechanically destabilized so that $x_i = \sigma_{th,i}-|\sigma_i| < 0$, it rearranges. Alternately, while $x_i > 0$, sites can spontaneously rearrange with an Arrhenius rate 
\begin{equation}
\lambda(x)=\omega_0 \exp[-\text{max}(x,0)^{\alpha}/T]. 
\label{eq:arrhenius}
\end{equation}
Here, the energy barrier for a site to thermally activate is taken to scale as $x^\alpha$ (cf. fig.~\ref{fig:schematic-failure}c), and the Boltzmann constant $k_B$ is absorbed into the definition of temperature. The scaling of saddle-node bifurcation suggests  $\alpha = 3/2$.   However, in other contexts, such as tissue development, it has been suggested that the potential energy barrier doesn't decay smoothly to zero, but is instead `cuspy', with $\alpha = 1$~\cite{popovic_inferring_2021}. For generality, and to test our scaling theories, we leave $\alpha$ free to vary. This activation rule has been used in several elastoplastic models to study steady-state rheology \cite{popovic2021thermalflow,ferrero_yielding_2021}, the transient approach to creep~\cite{popovic_scaling_2021},  relaxation dynamics~\cite{ferrero_relaxation_2014}, non-monotonic flow curves under mechanical vibration ($\alpha = 0$ and $\alpha =1$) \cite{le_goff_criticality_2019,goff_giant_2020}, the dynamics of super-cooled liquids (with no external stress, i.e. $\Sigma = 0$) \cite{ferrero_relaxation_2014,tahaei_scaling_2023}, and the size of avalanches \cite{korchinski_dynamic_2022}.

The system is externally driven in the strain-controlled protocol under simple shear, so that between plastic events, a site's stress grows as: $\sigma_i(t) = \dot{\gamma}t$. When a site $i$ rearranges at time $t_0$, it accrues a total increment of plastic strain $\delta\gamma_{pl,i} = \sigma_i(t_0)$. This strain increment is applied exponentially, so that $\gamma_{pl,i}(t-t_0) = \delta\gamma_{pl,i}(t_0) + (1-\exp(-(t-t_0)/\tau))\delta \gamma_{pl,i}$, where $\tau$ is a mechanical rearrangement timescale. We consider the case that $\tau = \omega_0^{-1}$, i.e. that  the plastic rearrangement timescale is of the same order of the thermal attempt frequency. Violating this assumption allows for interesting non-monotonic flow curves and critical fluctuations near the non-monotonic transient~\cite{goff_giant_2020}. We couple the sites with linear homogeneous isotropic elasticity. 
 \begin{equation}
    \tens{\gamma} = \frac{1}{2}\left[ (\vc{\nabla}\,\vc{u})+(\vc{\nabla}\,\vc{u})^T \right]
\end{equation}
and decompose the strain tensor into stress-free plastic and elastic parts as: $\tens{\gamma} = \tens{\gamma_{\text{pl}}} + \tens{\gamma_{\text{el}}}$. The elastic-strain contributes to the tensorial stress as:
\begin{equation} 
\tens{\sigma} = 2 \mu  \tens{\gamma_{\text{el}}} + \lambda \tr (\tens{\gamma_{\text{el}}})  \mathds{1} \,,
\end{equation}
where, for our simulations, we use $\mu = 1$ and $\lambda = \frac{1}{2}$. Numerically, we solve the elasticity partial differential equations using FEniCS~\cite{alnaes_fenics_2015,logg_dolfin_2010}.
We load our two-dimensional model under simple shear, with the displacement field prescribed on the boundaries as $\vc{u}(t) = (\gamma(t) \, y,0)^T$, leading to a uniform external loading on the $\sigma_{xy}$ component of the stress, with $\sigma_{xy}^{ext} =\dot{\gamma}t$. With simple shear, the dominant shear stress  at each site is $\sigma_{xy}$, making our model effectively scalar.  

At the mean-field level, we consider a thermal enhancement to the model of Hebraud and Lequeux~\cite{hebraud_mode-coupling_1998}. The HL model describes a distribution of local stresses that evolves with a biased diffusion, and rearrange when the local stress exceeds a yield threshold. 
\begin{equation}
    \partial_t P(\sigma,t) = D\partial_\sigma^2 P - \dot{\gamma}\partial_\sigma P +\Gamma\delta(\sigma) - \nu(\sigma) P 
\label{eq:HL_eqn}\end{equation}
The bias comes from external loading at a rate $\dot{\gamma}$, while the diffusion $D = a\Gamma$ occurs due to ``kicks'' from distant rearranging sites, occurring at a rate $\Gamma$.  Sites rearrange at a stress-dependent rate $\nu(\sigma)$. In the athermal model, this is taken to be at a rate $\omega_0$ when $|\sigma|>\sigma_{c}$, so $\nu(\sigma) = \Theta(|\sigma|-\sigma_c)$. The activity rate is then given by \begin{equation}\label{eq:HL_acivity_rate_eqn} \Gamma = \int \nu(\sigma)P(\sigma)\dup \sigma\,.\end{equation} This is coupled to the diffusion constant $D=a\Gamma$ by a scaling prefactor $a$. Since the number of sites is conserved, $\Gamma$ also enters as the reinjection rate of rejuvenated sites at $\sigma = 0$ through the $\Gamma\delta(\sigma)$ term. Equations~\ref{eq:HL_eqn} and~\ref{eq:HL_acivity_rate_eqn} describe the HL equations. The flow-stress at some point in time is then simply:
\begin{equation}
    \Sigma(t) = \int \sigma P(\sigma,t)\dup\sigma 
\end{equation}

When $a<\sigma_c^2/2$ the athermal HL model exhibits a Herschel-Bulkley average flow stress, with \begin{equation}\Sigma = \Sigma_c(\sigma_c,a) + \dot{\gamma}^{1/2} + \mathcal{O}(\dot{\gamma}^1)\,,\label{eq:HB}\end{equation} corresponding to a flow exponent $\beta=2$~\cite{agoritsas_relevance_2015}. For $a = \sigma_c^2/2$, the system acts as a power-law fluid, with $\Sigma\sim\dot{\gamma}^{1/5}$, and for $a > \sigma_c^2/2$ a Newtonian fluid $\Sigma\sim \dot{\gamma}$. For what follows, we will only consider $a<\sigma_c^2/2$. This model has been enriched in numerous ways, for instance including disorder~\cite{agoritsas_relevance_2015,agoritsas_non-trivial_2017}, fat-tailed stress kick distributions~\cite{lin_microscopic_2018}, oscillatory shear~\cite{parley_mean-field_2022}, and temperature~\cite{popovic_scaling_2021}. By studying the transient dynamics from a non-trivial $P_0(\sigma)$, creep~\cite{parley_aging_2020} and brittle yielding~\cite{barlow_ductile_2020,parley_towards_2023} can also be studied. To make contact with our EPM, we follow ref.~\cite{popovic_scaling_2021} and include a temperature-dependent yielding rate that allows sites to yield at $\sigma<\sigma_c$, as $\nu(\sigma) = \omega_0 \exp[ -\text{max}(\sigma_c-|\sigma|,0)^\alpha/T]$. Numerically and analytically, the strategy for solving these equations self-consistently  is done iteratively. A value for $D_0$ is guessed (at low strain-rates, $D = \frac{1}{2}a\dot{\gamma}$ is good to first order), and the corresponding HL equations are solved for fixed $D_0$, yielding $P_0(\sigma)$. $D_{i+1}$ is then computed, using $D_{i+1} = \int \nu(\sigma)P_i(\sigma)\dup\sigma$. Numerically, we solve the HL equation (eq.~\ref{eq:HL_eqn}) using finite-elements in one dimension using the FEniCS software package~\cite{alnaes_fenics_2015,logg_dolfin_2010}. For $\alpha=1$, we have an exact analytical solution, and numerical integration is not necessary.

\section{Continuous Flow}
\begin{figure}
    \centering
    \includegraphics[]{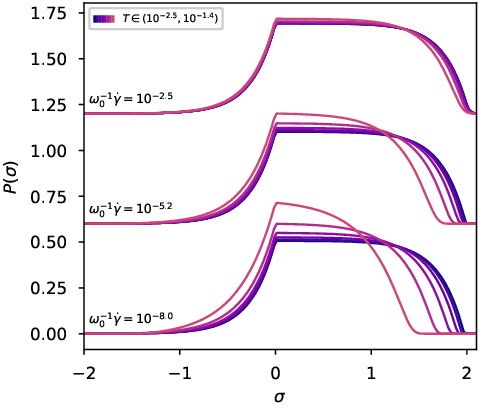}
    \caption{Exact solutions for the  $P(\sigma)$ distribution for different temperatures and driving rates (with an arbitrary offset for visual clarity) in the thermal HL model, for $a=0.5$, $\alpha =1$, and $\sigma_c = 2$. \label{fig:psigma_HL}}
\end{figure}
If a finite system is driven slowly enough, there is a separation of timescales between avalanches. As the driving rate or temperature is increased, this gives way to temporally overlapping flow, with multiple avalanches nucleating independently. Correspondingly, since stress cannot be dissipated before new avalanches are initiated, there is an increase in flow stress. Larger systems enter the continuous flow regime at lower temperatures and driving rates, and since fluctuations are sub-extensive and decorrelated, stress fluctuations shrink with system size. In the  $L\rightarrow \infty$ limit, all flow is continuous and smooth.  This is the mean-field limit we are considering with the HL model.

We will seek steady-state solutions to the HL equations, which can be obtained exactly for $\alpha = 1$, and numerically for $\alpha>1$. For the case $\alpha = 1$, the solution in the range $\sigma\in (0,\sigma_c)$ was essentially worked out in ref.~\cite{popovic2021thermalflow}, and we build on this prior work in  appendix~\ref{sec:HL_sln}, where we derive the exact solution for $\alpha = 1$ for all $\sigma$. Such exact solutions are displayed in Fig.~\ref{fig:psigma_HL}, for various driving rates and temperatures.

\begin{figure*}
    \centering
    \includegraphics[]{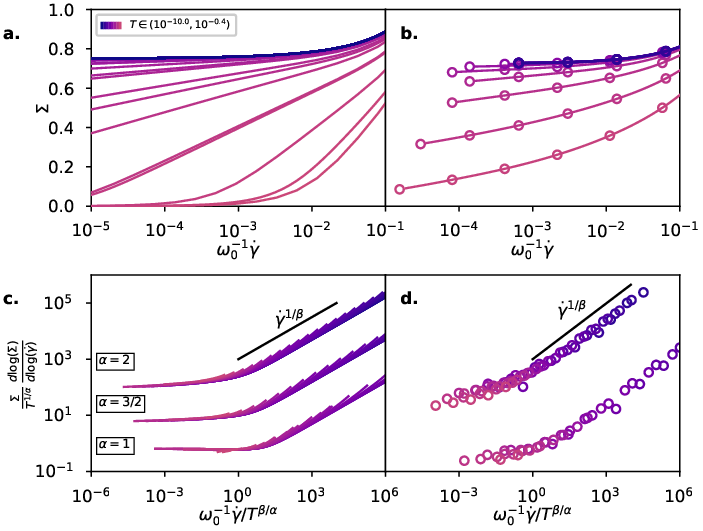}
    \caption{ \textbf{a. and b.} Flow curves for different temperatures with $\alpha=1$ for the HL model (a) at $a=0.5$ and in a corresponding EPM simulation (b). Solid lines between EPM data point are cubic spline interpolations to ease identification of different temperatures. \textbf{c. and d.} $\dd{\log(\Sigma)}{\log(\dot{\gamma})}$, re-scaled to effect a collapse. $\beta=2$ for HL (c) and $\beta =\betaEPM$ for EPM (L=64) (d). Data are shifted vertically by $10^3$ for $\alpha =2$ and $10^{\betaEPM}$ for $\alpha = 3/2$ for clarity.  Data with $\Sigma < 0.05$ are omitted for visual clarity, because they do not fall onto the collapse curve (see text).}
    \label{fig:flowcurve_continuous}
\end{figure*}
It's worth making a few observations at this point: At low temperatures, there is a boundary layer close to $\sigma_c$, where sites evaporate. As the temperature increases, sites become thermally activated at $\sigma < \sigma_c$. This effect is most pronounced at low driving rates, and essentially non-existent for high driving rates and low-temperatures.

This suggests that there is a rheological crossover at some temperature-dependent driving rate  $\dot{\gamma}_{cross}(T)$. Indeed, by studying the average flow-stress $\Sigma(T,\dot{\gamma}) = \int P(\sigma)\sigma \dup \sigma$ (cf. Fig.~\ref{fig:flowcurve_continuous}), we see that for cold systems at high-driving, we recover the HB behaviour, while for higher temperatures, there is an essentially logarithmic dependence on driving rate. This manifests in both the HL data (cf. Fig.~\ref{fig:flowcurve_continuous}a) as well as the corresponding EPM data (cf. Fig.~\ref{fig:flowcurve_continuous}b).
For $\alpha = 1$, we can obtain this result analytically, finding asymptotically (for small $a$ and $\dot{\gamma}$)  \begin{equation}\Sigma = \frac{1}{2}\left(\sigma_c - T\log\left[\frac{4T^2}{a\omega_0^{-1}\dot{\gamma}}\right]\right)\label{eq:sigma_asymptotic_alpha_1_HL}\end{equation} 
for $T^{\beta} > \omega_0^{-1}\dot{\gamma}$.  This logarithmic dependence shows up naturally as straight lines on the semi-log plots of fig.~\ref{fig:flowcurve_continuous}a and b.  %

We test the prediction between athermal and thermal flow , $\dot{\gamma}_{cross}\sim T^{\beta/\alpha},$ in fig.~\ref{fig:flowcurve_continuous}c and d.  We consider the logarithmic derivative of stress suggested in ref.~\cite{yu_active_2020}. This quantity has the benefit of scaling differently in the HB and logarithmic regimes, while not requiring a delicate fitting procedure to obtain $\Sigma_c$ and $1/\beta$, which would be required to collapse $\Sigma-\Sigma_c \sim T^{1/\alpha}f(\dot{\gamma}/T^{\beta/\alpha})$ directly. In the HB regime, with $\Sigma = \Sigma_c + A\dot{\gamma}^{1/\beta}$, the logarithmic derivative scales as $\dd{\log(\Sigma)}{\log(\dot{\gamma})} \sim \frac{\dot{\gamma}^{1/\beta}}{\beta \Sigma} $. Meanwhile, in the logarithmic regime, we expect the scaling $\dd{\log(\Sigma)}{\log(\dot{\gamma})} \sim \frac{T^{1/\alpha}}{\Sigma}$. Our predicted rescaling serves to collapse both our HL data and our 2d EPM data in the continuous flow regime (cf. fig.~\ref{fig:flowcurve_continuous}c and d). To effect a good collapse, however, we do need to restrict ourselves to the data from the logarithmic and HB regimes. Obviously, the logarithmic scaling predicted by eq.~\eqref{eq:sigma_asymptotic_alpha_1_HL} fails when the $T\log(4T^2/a\omega_0^{-1}\dot{\gamma})$ term is of order $\sigma_c$ and therefore the stress $\Sigma\ll1$, so we consider only data with $\Sigma>0.05$ for the collapse. In the ultra-low stress limit, we find $\Sigma\sim\dot{\gamma}$, and discuss this further in appendix~\ref{sec:HL_sln}. Although the EPM has features not present in the HL model, namely long-tailed kick distributions and a finite dimension, the main features of HL seem to survive for the 2d model: namely the crossover scaling as $\dot{\gamma}_{cross}\sim T^{\beta/\alpha}$ (albeit with different values of $\beta$ for 2d and HL) and the essentially logarithmic rheology for $\dot{\gamma} < \dot{\gamma}_{cross}$. 

This data may appear to be in tension with the $\dot{\gamma}\sim T^{1/\alpha}$ crossover scaling we predicted in ref.~\cite{korchinski_dynamic_2022}. There, the data supported a crossover at $\dot{\gamma}_{cross} \sim T^{1/\alpha}$. Here, we are being careful to restrict our analysis to sufficiently large, hot, and quickly driven systems so that flow is continuous. For this case of continuous flow, it appears that the rheological crossover between athermal and thermal flow indeed scales at $T^{\beta/\alpha}$. The $\beta$ exponent enters because of cooperativity between temporally overlapping flows. The situation is somewhat different in the intermittent regime, which will be discussed in the next section.

\section{Intermittent Flow}

\subsection{Activation threshold $x_a(T,\dot{\gamma})$}
For sufficiently small, cold, and slowly driven systems \cite{korchinski_dynamic_2022}, there is a separation of timescales between periods of cascading plastic activity. The duration of an avalanche is typically much shorter than the time to trigger the next one. In steady state, the amount of stress loaded into the system between two avalanches must, on average at least, equal the energy dissipated by an avalanche. In the absence of thermal activation, it is the weakest site of the system, capable of withstanding an additional stress $\xm$, that controls how much stress can be loaded  before a new avalanche is initiated. Hence, \begin{equation} \langle \Delta \Sigma\rangle_{av} = \xmavg\end{equation}  in athermal systems. Moreover, avalanches are power-law distributed with a system-size dependent truncation, i.e. $p(s)\sim s^{-\tau}g(s/L^{d_f})$, yielding $\langle s\rangle \sim L^{d_f(2-\tau)}$. If the distribution of stable sites in the system exhibits a pseudogap, $p(x)\sim x^{\theta}$, then extreme value statistics give $\xmavg\sim L^{-d/(1+\theta)}$, which gives the scaling relation \cite{lin_scaling_2014}
\begin{equation}
    \tau = 2-\frac{\theta d}{d_f(1+\theta)} \,. 
\end{equation}
There are additional finite-size corrections to $p(x\ll 1)$, namely  a shallower power law below residual stresses scaling as $L^{d_f-d}$ and the appearance of a terminal plateau below $L^{-d}$ that lead to a modified scaling relation \cite{korchinski_signatures_2021},
\begin{equation}
    \tau = 2-\frac{(d_f-d)\theta}{d_f} \,. \label{eq:tau_scaling_law_modified}
\end{equation}

When thermal activation is allowed to trigger otherwise mechanically stable sites, the amount of loaded stress can change in two ways: (i) the initiating site can fail with residual stability $x>0$, shortening the loading period, and (ii) thermal activation is stochastic, so it is not always the least stable site that yields first. 

\begin{figure}
    \centering
    \includegraphics{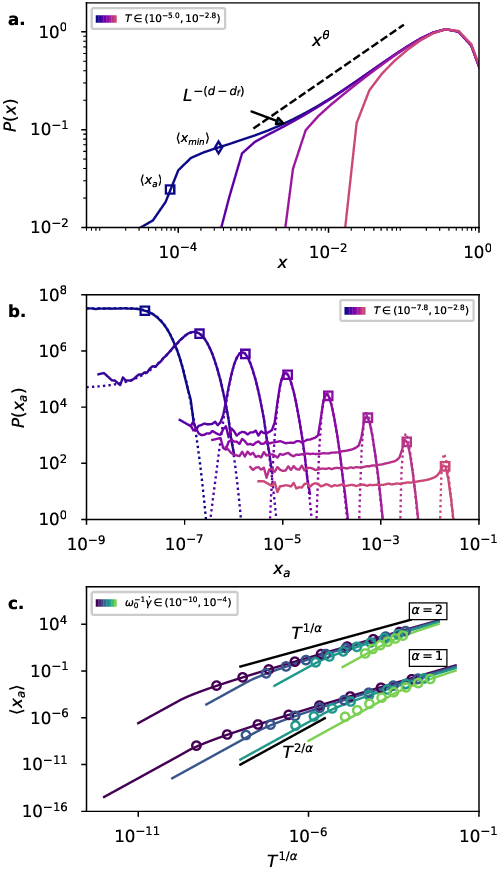}
    \caption{\textbf{a.} Distribution of stable sites after avalanches in the intermittent regime, $\alpha=1$, and $\omega_0^{-1}\dot{\gamma} = 10^{-8}$. $\xmavg$, $\langle x_a\rangle$ and the plateau crossover $x \approx L^{-(d-d_f)}$ are indicated for the coldest simulation. \textbf{b.} Stability distribution for the first activated site during an avalanche for $\alpha=1$ and $\omega_0^{-1}\dot{\gamma} = 10^{-8}$. Solid lines are data gathered in EPM, while dotted lines are theoretical curves. Squares indicate the average $\xaavg$.
    \textbf{c.} Average activation stabilities obtained from EPM simulations at different driving rates and temperatures, and for $\alpha=1$ (bottom) and $\alpha=2$ (top, shifted vertically by $10^6$ for visual clarity). Solid theoretical curves correspond to eq.~\eqref{eq:xaavg} for $\omega_0 T^{1/\alpha} >\dot{\gamma}$, while for lower temperatures we use $\xaavg\sim \omega_0 T^{2/\alpha}/\dot{\gamma}$. All data is taken for $L=256$ EPM.}
    \label{fig:pxa_xaavg}
\end{figure}

We focus on this first point in our subsequent analysis, but studying the activation scale $x_a(T,\dot{\gamma})$, which is the typical threshold for a stable site to yield via thermal activation. This is a key quantity to understand the effect of temperature on avalanche statistics and global rheology in the intermittent regime. Figure~\ref{fig:pxa_xaavg}a reveals a direct physical interpretation: below the typical value $\langle x_a(T,\dot{\gamma})\rangle$, the distribution of residual stability $P(x)$ becomes gapped as sites below  $\langle x_a(T,\dot{\gamma})\rangle$ are depleted. The size of the gap grows with increasing temperature.

We can compute $\langle x_a\rangle$ by considering the fate of a single site with some initial stability $x_0$ being driven towards instability at rate $\dot{\gamma}$, so $x(t) = x_0 - \dot{\gamma}t$. From eq.~\eqref{eq:arrhenius} when the site reaches the thermal stress scale $x_c=T^{1/\alpha}$, it should yield on a timescale $\approx 1/\omega_0$, suggesting an activation scale $x_a(T) = T^{1/\alpha}$. This is the approximation used in ref.~\cite{korchinski_dynamic_2022}, but it does not account at all for different driving rates. If the driving rate is sufficiently slow, then  the site could yield at $x>T^{1/\alpha}$. 
To capture this effect to first order, the activation rate $\lambda(x)$ is approximately constant for a time window $\Delta t \approx T/\dot{\gamma}$. Thermal activation occurs during such a time-widow if $\lambda(x_a) \Delta t \approx 1$. This occurs at an activation scale $x_a = \left[T\log\left(\omega_0T/\dot{\gamma}\right)\right]^{1/\alpha}$, which we will see is essentially correct. We can obtain a more precise result by computing the distribution of activation stabilities for a single site $p(x_a|x_0)$. The probability that a site survives for time $t$ without activating is $P(t_{fail} > t) = \exp[-\int_{0}^{t} \lambda(x(t'))  dt'$, and so the probability of failing at exactly time $t$ is $P(t_{fail} = t) = P(t_{fail}>t)\lambda(x(t))$. With a change of variables $x = x_0 -\dot{\gamma}t$, we have
\begin{equation}
    P(x_a|x_0) = \frac{1}{\dot{\gamma}}\lambda(x_a)\exp\left[-\frac{1}{\dot{\gamma}}\int^{x_0}_{x_a} \lambda(x') dx'\right]\,.
\end{equation}
The first site to fail will usually be starting from $x_0 = \xm$. However, the rate of failure for sites with $x>\xm$ is low, and in practice if we use $x_0 \rightarrow \infty$, we  find excellent agreement with $P(x_a)$ collected numerically (cf. fig.~\ref{fig:pxa_xaavg}b), and $p(x_a)$ takes a simple analytical form, developed in appendix~\ref{sec:pxa_discussion}.

We can use $P(x_a)$ to compute $\xaavg$. Because the thermal activations functions we consider are Arrhenius, we have an exponential of an exponential, and $P(x_a|x_0)$ is sharply peaked (so long as $T \omega_0/\dot{\gamma} \gg 1$) as can be seen in Fig.~\ref{fig:pxa_xaavg}b. So, we  approximate $\xaavg = \int x_aP(x_a)dx_a \approx \argmax( P(x_a) ) $. This is conveniently computed by setting $0 = \pp{}{x_a}\log(P(x_a))$, which leads to $\frac{\omega_0}{\dot{\gamma}}\exp\left[ -x_a^{\alpha}/T \right] = \alpha x_a^{\alpha-1}/T $. For $\alpha = 1$, this gives $\xaavg = T\log(\omega_0T/\dot{\gamma})$, consistent with our initial prediction. For $\alpha > 1$, the general solution is
\begin{equation}
    \xaavg = \left[ \tfrac{\alpha-1}{\alpha} T \lambW\left( \tfrac{1}{\alpha-1}\left[ \frac{T}{(\alpha\dot{\gamma}/\omega_0)^\alpha} \right]^{\frac{1}{\alpha-1}} \right)  \right]^{1/\alpha} \label{eq:xaavg}
\end{equation}
where $\lambW$ is the Lambert-W function, defined by $z = \lambW(z)e^{\lambW(z)}$.

There is a change in behaviour when $\dot{\gamma} \approx \omega_0 T^{1/\alpha}$, i.e. precisely when the assumption that $P(x_a)$ is sharply peaked breaks down, and $P(x_a)$ becomes flat. For colder temperatures, it becomes likely that a site starting at $x>0$ will reach $x=0$ without being thermal activated. In this regime, we approximate the activation function  as step function at $x_c = T^{1/\alpha}$ so that $\lambda(x) =\omega_0\Theta(x_c - x)$, so $P(x_a) \approx \frac{\omega_0}{\dot{\gamma}}\exp\left[-\frac{x_c-x}{\dot{\gamma}/\omega_0}\right]\Theta(x_c-x)$. Hence, $\xaavg = \dot{\gamma}/\omega_0 \left(e^{-{x_c\omega_0/\dot{\gamma}}}-1\right)+x_c$ and when $x_c=T^{1/\alpha} \ll \dot{\gamma}/\omega_0$, we find $\xaavg \approx \omega_0T^{2/\alpha}/(2\dot{\gamma})$. 

We verify these findings in numerical simulations of our elastoplastic model in fig.~\ref{fig:pxa_xaavg}c, and find excellent agreement between our prediction for $\xaavg$ and simulation data at $L=256$. Agreement is excellent at smaller system sizes as well, though we do not include this data to reduce visual clutter.

\begin{figure*}
    \centering
    \includegraphics[width=\linewidth]{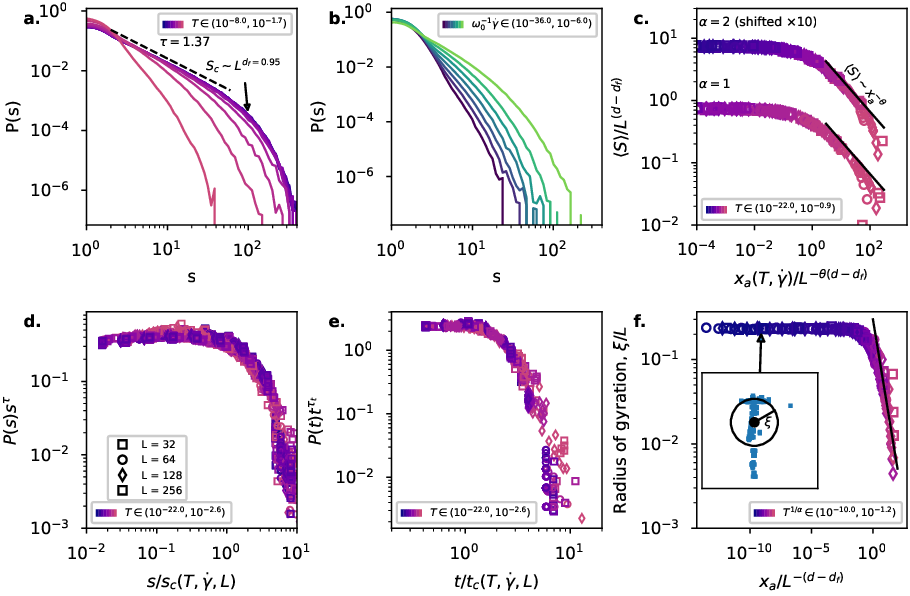}
    \caption{Avalanche properties for systems at different temperatures, driving rates, $\alpha=1$ and $\alpha=2$, and system size. \textbf{a.} Avalanche size distribution, for $L=128$, $\omega_0^{-1}\dot{\gamma} = 10^{-6}$, and $\alpha=1$. \textbf{b.} Avalanche size distribution, $L=128$, $T=4\times 10^{-3}$, and $\alpha=1$. \textbf{c.} Average avalanche size, for systems at different driving rates and temperatures collapsed onto a universal curve. \textbf{d.} Avalanche size distribution, varying temperature, driving rate, $\alpha$, and system size, collapsed onto a universal curve. \textbf{e.} Avalanche duration distribution, collapsed onto a universal curve. \textbf{f.} The radius of gyration collapsed onto a universal curve. Solid line corresponds to our prediction $\nu = \frac{\theta}{d_f(2-\tau)}$ . Inset depicts how the radius of gyration for a single avalanche relates to its geometric extent. }
    \label{fig:avalanches}
\end{figure*}

\subsection{Avalanches}
Next, we can turn to computing properties of avalanches in the intermittent regime. As can be seen in fig.~\ref{fig:avalanches}a, increasing temperature (up to a system-size dependent cutoff) truncates the scale-free avalanche distribution. Meanwhile, increased driving rate prevents thermal activation and decreases the activation scale $x_a$, increasing avalanche size (cf. Fig.~\ref{fig:avalanches}b).  

With $\xaavg$ now under control, we can compute the expected avalanche size as $\langle s\rangle L^{-d} = \langle \xm - x_a\rangle$. 
The usual extreme value argument applies that fixes $\xm$: \begin{equation}L^{-d} \sim \int_0^\xmavg p(x)\dup x\,. \label{eq:xmavg_extreme_value_argument}\end{equation} 
The $p(x)$ stability distribution in athermal finite systems diverges from the pseudogap scaling $p(x)\sim x^\theta$ below a stress scale $x<L^{-(d-d_f)}$ induced by the largest avalanches~\cite{korchinski_signatures_2021}. Below this scale, $p(x)$ varies slowly, and for large system sizes approaches a plateau.  Non-zero temperatures cause $p(x)$ to become gapped below the scale $x_a$. If $x_a > L^{-(d-d_f)}$, then 
\begin{equation} \xmavg = \left(x_a^{1+\theta} + A L^{-d} \right)^{1/(1+\theta)}\,,\label{eq:xmavg_above_plateau}\end{equation} for some constant $A$. Since $x_a > L^{-d/(1+\theta)}$ (as we assumed  $x_a>L^{-(d-d_f)}$ and $d-d_f\approx 1.05  < d/(1+\theta)\approx 1.31$ for $d=2$, 
we can expand eq.~\eqref{eq:xmavg_above_plateau} to first order and obtain
$\xmavg - x_a \sim L^{-d}x_a^{\theta}$, 
which implies that the average avalanche size scales as \begin{equation}
\langle s\rangle \sim x_a^\theta  \,. \label{eq:s_avg_T_scaling_regime}
\end{equation}

However, if both $\xmavg$ and $x_a$ are in the finite-size dependent plateau
(for which $p(x\ll L^{-(d-d_f)} \sim L^{-p=\theta(d-d_f)}$) dominates the scaling), eq.~\eqref{eq:xmavg_extreme_value_argument} yields $L^{-d} \sim L^{-p}(\xmavg - x_a)$ implying that for low enough temperatures, 
\begin{equation}
    \langle s\rangle \sim L^{-p} = L^{\theta(d-d_f)}\,. \label{eq:s_avg_L_scaling_regime}
\end{equation}
We test the scaling collapse in fig.~\ref{fig:avalanches}c for simulations with varied driving rate, system size ($L=32$ to $L=256$), and temperature. For both $\alpha = 1$ and $\alpha = 2$, we confirm that our scaling prediction eq.~\eqref{eq:s_avg_T_scaling_regime} applies for high temperatures, when $x_a> L^{-\theta(d-d_f)}$ and that eq.~\eqref{eq:s_avg_L_scaling_regime} applies at low temperatures, indicating avalanches are truncated by finite system size. 

Avalanches are scale free, up to a cutoff $s_c(x_a,L)$, so that $p(s)\sim s^{-\tau}G(s/s_c)$, where $G(z)$ is a scaling function satisfying $G(z \ll 1)\approx 1$ and $G(z>1) \rightarrow 0$.  This scaling ansatz yields implies that the cutoff controls the average avalanche size, with $\langle s\rangle = s_c^{2-\tau}$.  So, our calculation for the average avalanche size gives us a handle on the largest avalanches in the system by inverting the relation, with $s_c\sim \langle s\rangle^{1/(2-\tau)}$. The cutoff therefore scales as 
\begin{equation}
    s_c\sim x_a^{-\theta /(2-\tau)} f\left( x_a/L^{-(d-d_f)} \right)\,,\label{eq:s_c_in_xa}
\end{equation}
where $f(z)$ is a scaling function that goes as $f(z\ll 1)\sim z^{\theta/(2-\tau)}$ and $f(z>1)\approx 1$. Note, that since $s_c\sim L^{d_f}$ in the athermal limit, this scaling form implies that $L^{\theta(d-d_f)/(2-\tau)} \sim L^{d_f}$ implying $\tau = 2-\frac{(d_f-d)\theta}{d_f}$, i.e. recovering eq.~\eqref{eq:tau_scaling_law_modified}.
In fig.~\ref{fig:avalanches}d, we effect a satisfactory scaling collapse using the  functional form 
$f(x) = 1/\sqrt{1+Bx^{-2(\theta/(2-\tau))}}$, fitting $B$ to the $\langle s\rangle $ data. 

The theory of critical phenomena suggests that scale free physics is a result of a diverging correlation length. Consequently, avalanches of a physical extent $\ell$ have a fractal scaling for size as $s\sim \ell^{d_f}$ and for duration  as $t\sim \ell^z$. For finite systems, the diverging correlation length reaches that of the system, giving the usual cutoffs $s_c\sim L^{d_f}$ and a maximum duration $t_c\sim L^{z}$. Hence, we expect that, on average, avalanche durations are related to their sizes by $t \sim s^{z/d_f}$, and that the duration cutoff should simply scale as $t_c\sim s_c^{z/d_f}$. We measure avalanche durations (pruning short duration avalanches with $t < 3\omega_0^{-1}$ that fall outside the scaling regime for clarity), and obtain a good collapse across temperature, driving rate, $\alpha$, and system size in fig.~\ref{fig:avalanches}e. 

In fig.~\ref{fig:avalanches}f, we directly confirm the truncated correlation length picture by measuring the radius of gyration for avalanches, and using these geometric measures of avalanche extent to infer the correlation length. 
The radius of gyration for a single avalanche comprised of elementary events at locations $\vec{r}_i$ is defined as 
\begin{equation}
    R_g^2 \equiv \frac{1}{N}\sum_i^N (\vec{r}_i - \vec{r}_0)^2\,,
\end{equation}
where $\vec{r}_0 = \frac{1}{N}\sum_i^N \vec{r}_i$ is the center of mass.  We define the correlation length for a collection of avalanches (indexed by $j$) as it is done in percolation~\cite{grimmett_percolation_1999}, 
\begin{equation}
    \xi^2 \equiv \frac{\sum_j R_g(j)^2 s_j^2}{\sum_j s_j^2}
\end{equation}
In the criticality picture, we expect 
\begin{equation}\xi^{d_f} \sim s_c\sim x_a^{-\theta/(2-\tau)} \,. \end{equation}
We find that all correlation lengths fall onto the same universal curve, with $\xi \sim L$ when $x_a < L^{-(d_f-d)}$ and $\xi \sim x_a^{\frac{-\theta}{d_f(2-\tau)}}$ in fig.~\ref{fig:avalanches}f. We summarize the exponents used to effect curve collapses in Table.~\ref{table:exponents}

\begin{table}
\caption{A table of exponents used to effect scaling collapses, along with their definitions and relevant scaling relations. The two values for $\beta$ refer to the 2d EPM vs HL model, respectively. Starred $({}^*)$ values are inferred from the other exponents.
\label{table:exponents}}
    \begin{tabular}{c|c|c}
        Exponent & Value & Definition  \\ \hline \hline 
        $\theta$ & 0.57 & $p(x)\sim x^{\theta}$ for $x>L^{-(d-d_f)}$ \\ \hline  
        $\tau$ & 1.37 & $p(s)\sim s^{-\tau} g(s/s_c)$  \\ \hline 
        $d_f$ & 0.95 & $s \sim \ell^{d_f}$ and $s_c\sim L^{d_f}$ in AQS  \\ \hline 
        $\tau_t$ & 2.0 & $p(t) \sim t^{-\tau_t} f(t/t_c)$  \\ \hline 
        $z$ & 0.6 & $t \sim \ell^{z}$ or $t_c\sim L^{z}$ in AQS  \\ \hline 
        $\beta$ & \betaEPM/2 & $\dot{\gamma}\sim \Delta\Sigma^\beta$  \\ \hline 
        $\nu$& $0.89^*$ & $\xi \sim \Delta\Sigma^{-\nu}$ \\ 
         & & $\nu = \theta / (d_f(2-\tau)) = 1/(d-d_f)$  \\ \hline
        $1/\sigma$ & & $s_c\sim \Delta\Sigma^{-1/\sigma}$   \\
         & 0.9$5^*$ & $1/\sigma = \nu d_f$ and $\sigma\theta = 2-\tau$  \\ 
    \end{tabular}
\end{table}

\subsection{Flow stress}
\begin{figure*}
    \centering
    \includegraphics[width=\linewidth]{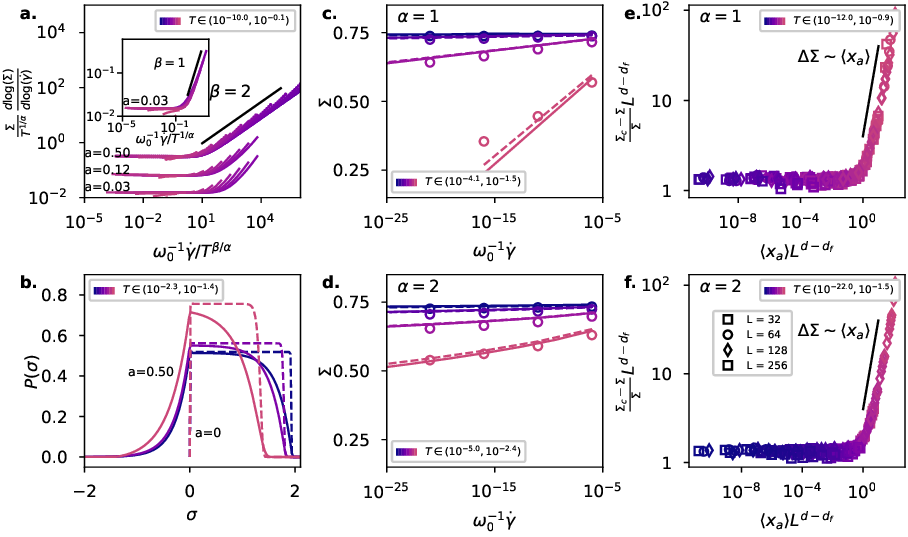}
    \caption{\textbf{a.} Rheological transition as a function of $a$ for the thermal HL model ($\alpha = 1$), rescaled to effect a collapse for large $a$. Small values of $a$ do not collapse. Inset: rheological collapse, rescaled for $\beta = 1$, for the $a=0.03$ HL solution. \textbf{b.} The $p(\sigma)$ solutions to the thermal HL model for $a=0.5$ (solid) and $a=0$ (dashed), for different temperatures at a driving rate $\omega_0^{-1}\dot{\gamma} = 10^{-8}$. \textbf{c. and d.} Rheology in the intermittent regime for both $\alpha=1$ (c) and $\alpha = 2$ for $L=128$ EPM data. (d). Circles are $L=128$ EPM data, dashed lines are fitted results from eq.~\ref{eq:stress_response_intermittent}, while solid lines correspond to $a=0$ in the thermal HL model. \textbf{e. and f.} Finite-size scaling for the transition between $L$ and $\xaavg$ control of  distance to $\Delta \Sigma$. }
    \label{fig:intermittent_rheology}
\end{figure*}

In the previous section, we described how fluctuations in stress, i.e.~avalanches, are affected by temperature and driving rate. Next, we will work out the contribution of temperature and driving rate to the change in macroscopic flow stress, $\Sigma(T,\dot{\gamma}) = \int \sigma P(\sigma)\dup\sigma$. Given the central role of $x_a$ in describing avalanche sizes, it is reasonable to suspect that $x_a$, being a stress scale, will also govern the macroscopic stress gap as $\Sigma_c - \Sigma(T,\dot{\gamma}) \propto x_a$. As we will see, this is essentially correct. However, this guess sits in tension with the rheology of continuous flow. In continuous flow, there is a crossover between temperature and driving rate effects at $\dot{\gamma} \sim T^{\beta/\alpha}$, while $x_a$ sees a crossover when  $\dot{\gamma} \sim T^{1/\alpha}$. 

In continuous flow, $\beta$ enters as $(\Sigma-\Sigma_c)^\beta \sim \dot{\gamma}$ with $\beta>1$. This means a small increase in stress leads to a super-linear increase in flow. The introduction of more stress leads additional sites to fail, leading to a greater stress diffusion constant $D$, which in turn leads to additional failing sites. This positive feedback from cooperating sites necessitates $\beta>1$. However, in the intermittent regime, avalanches are isolated events that do not overlap temporally, and therefore do not cooperate.

To capture this lack of cooperativity and  make contact between the intermittent and continuous flow regimes, we can adapt the thermal HL model to the study of non-overlapping rheology by considering the $a\rightarrow 0^+$ limit.  Physically, this would correspond to taking a sufficiently large system, with $L\gg \xi$, so that avalanches are controlled by temperature and driving rate effects, but not so large that avalanches temporally overlap. Then, even the largest avalanches are small with regard to the whole system.  We could imagine coarse-graining the system to the level of blocks of size $\xi$ and therefore to independent nucleation events. In this limit, we imagine that only one region is yielding at a given moment, so there are no cooperative effects between avalanches.  In this limit, as $a\rightarrow 0$, the rheological transition between HB flow and logarithmic flow no longer occurs at $\dot{\gamma} \sim T^{\beta/\alpha}$, but instead at $T^{1/\alpha}$ (cf. Fig.~\ref{fig:intermittent_rheology}a and associated inset).  Consistently, the HB flow for $a\rightarrow 0$ is shallower, with $\Sigma-\Sigma_c\sim \dot{\gamma}$ suggesting $\beta\rightarrow 1$ for $a=0$, which can be directly observed in the inset of Fig.~\ref{fig:intermittent_rheology}a. 

Setting the activity-diffusion prefactor $a$ to zero results in a simplified set of HL equations, 
\begin{equation}
0=\partial_t P = -\dot{\gamma}\partial_\sigma P - \nu(\sigma)P\,,
\end{equation}
where $\nu(\sigma) = \omega_0^{-1}\exp(-(\sigma_c-\sigma)^{\alpha}/T)$.  Being first order, these equations can be solved exactly for any $\alpha$, the full treatment for which is given in appendix~\ref{sec:niHL_sln}. The salient features of the reduced model are as follows: $\beta = 1$, which immediately leads to a rheological crossover scaling as $T^{1/\alpha} \sim \omega_0^{-1}\dot{\gamma}$, and a boundary layer occurring at $\sigma_a = \sigma_c - x_a$, where the activation threshold $x_a$ is identical to that given by eq.~\eqref{eq:xaavg} from the biased random walker picture. Solutions to the thermal HL equations with $a>0$ and $a=0$ are compared in Fig.~\ref{fig:intermittent_rheology}b.
Both models agree as to where the extinction of sites should occur, with $P(\sigma)$ vanishing for $\sigma > \sigma_c-x_a$, but the boundary layer for the $a>0$ case is much wider due to the presence of diffusion. 

Finally, we can compute the expected decrease in flow stress due to premature activation of sites. For the $a=0$ model, $P(\Sigma)$ is essentially constant below $x_a$, suggesting that $\Delta\Sigma = \Sigma_c-\Sigma(T,\dot{\gamma}) \sim x_a$. Therefore, we expect more generically that, 
\begin{equation}\Sigma(T,\dot{\gamma}) = \Sigma_c - K x_a\,,\label{eq:stress_response_intermittent}\end{equation}
for some material-dependent prefactor $K$. This of course works well for the thermal HL model with $a\rightarrow 0$, but also for our thermal EPM data taken in the intermittent regime.  We find excellent agreement with this prediction at all but the highest temperature ranges for both $\alpha=1$ (cf.~fig.~\ref{fig:intermittent_rheology}c) and $\alpha=2$ (cf.~fig.~\ref{fig:intermittent_rheology}d), fitting only the parameters $K$ and $\Sigma_c$ to all our data sets. For finite size systems, $\Sigma_c-\Sigma \sim \xaavg$ only when $\xaavg > x_{a,cross}(L) \sim L^{-(d-d_f)}$. For $x_a < x_{a,cross}$, the flow stress is fixed by the system size, which then predicts $\Sigma_c - \Sigma =  \sigma_{a,cross} \sim L^{-(d-d_f)}$, which we confirm by the collapse of Figs.~\ref{fig:intermittent_rheology}e and f. 

The rheology of intermittent flow shares some features with that of continuous flow. In both cases, increased driving rate or decreased temperature increase the flow stress. Additionally, $\alpha > 1$ introduces a subtle curvature to the semilog rheology curves. In the intermittent flow, this originates from the second-order (and higher) terms coming from the Lambert-W function, i.e. $x_a \sim [T \lambW(z)]^{1/\alpha} \sim [T(\ln(z)-\ln(\ln(z))+\ldots)]^{1/\alpha}$. However, between continuous and intermittent flow, driving rate and temperature compete differently. In continuous flow, the overlapping of avalanches mean that the $\beta$ exponent enters into the crossover between temperature and strain rate at $T^{\beta/\alpha} \sim \omega_0^{-1}\dot{\gamma}$. Meanwhile, for intermittent flow, the strain-rate and temperature competition crossover occurs at $T^{1/\alpha }\sim \omega_0^{-1}\dot{\gamma}$. 

Lastly, identifying $\Sigma_c-\Sigma = \Delta\Sigma \sim x_a$ implies that our scaling relations for avalanche size in terms of $x_a$ can apply to a  more general set of relations. In particular, $s_c \sim \Delta\Sigma^{-1/\sigma}$ taken with eq.~\eqref{eq:s_c_in_xa} implies the scaling relation $\sigma\theta = 2-\tau$. Since $1/(\sigma\nu) = d_f$, 
this implies $\nu = \frac{\theta}{d_f(2-\tau)}$. For our measured exponents, this predicts $\nu = 0.89$, somewhat lower than what has been measured in ref.~\cite{jocteur_yielding_2024} where they find $\nu = 1.11$. This discrepancy could indicate that a stress gap opened by finite temperature behaves differently than a stress-gap opened at zero temperature (where triggering is always extremal).

\subsection{Discussion}
\begin{figure}
    \centering
    \includegraphics[width=\linewidth]{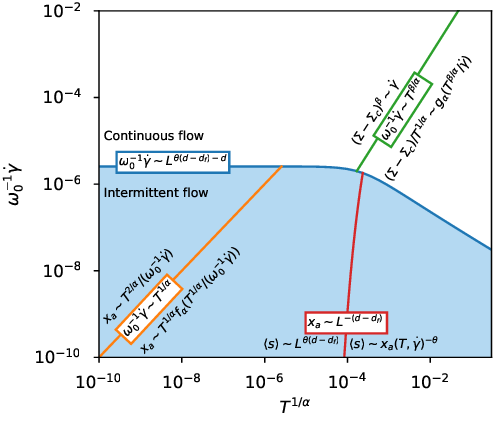}
    \caption{Schematic dynamical phase diagram for the amorphous yielding transition at finite temperature, driving rate, and system size. This phase diagram is generated to match EPM data for $L=512$.}
    \label{fig:phase_diagram}
\end{figure}
We summarize the different dynamical phases of thermal amorphous yielding in Fig.~\ref{fig:phase_diagram}. Our main finding is that the continuous and intermittent regimes have different rheology. We delineate these regimes by comparing the average loading time between avalanches to the mechanical rearrangement timescale $\tau_0$. Since we take the mechanical rearrangement time to be the inverse of the Arrhenius attempt frequency $\tau_0 = \omega_0^{-1}$, the dividing line between continuous and intermittent flow occurs when $\omega_0^{-1}= \langle t_{load} \rangle= \frac{1}{\dot{\gamma}} \langle s\rangle L^{-d}$.  In the limit of low temperature, the average avalanche size, this gives a crossover driving rate that scales with the system size as $\omega_0^{-1}\dot{\gamma} \sim L^{\theta(d-d_f)-d}$.  However, when $x_a \gg L^{-(d-d_f)}$, $x_a$ is dominated by temperature effects and the crossover driving rate goes as $\omega_0^{-1}\dot{\gamma} =  x_a(T,\dot{\gamma})^{-\theta}L^{-d} \approx  T^{-\theta/\alpha}L^{-d}\log\left(\frac{T}{(\alpha \dot\gamma / \omega_0)^\alpha}\right)^{1/\alpha} $.

In the intermittent regime, the rheology is controlled by the activation stress scale $\Delta\Sigma\sim x_a$. The scaling for $x_a$ changes when thermal activation becomes the dominant form of excitation, i.e. when $\omega_0^{-1}\dot{\gamma} < T^{1/\alpha}$. In this regime, $\Delta \Sigma \sim T^{1/\alpha} f_\alpha(T^{1/\alpha}/(\omega_0^{-1}\dot{\gamma}))$. 

In contrast, in the continuous regime, the rheology has a Herschel-Bulkley character: an increase in stress or temperature leads to a nonlinear increase in further activity characterized by the flow exponent $\beta$. This can only occur because flow consists of temporally overlapping avalanches. These avalanches reinforce each other by contributing overlapping long-ranged mechanical noise. In this case, the flow behaves as $\Delta \Sigma \sim T^{1/\alpha}g_\alpha(T^{\beta/\alpha}/(\omega_0^{-1}\dot{\gamma}))\,.$ 

In the continuous regime, the correlation length shrinks as the driving rate is increased, since $\xi \sim \Delta \Sigma^{-\nu}\sim \dot{\gamma}^{\beta/\nu}$. Conversely, in the intermittent regime,  increased driving rate grows the correlation length. This is because increased driving rate shrinks the effect of temperature, and brings $\Sigma\rightarrow \Sigma_c^{-}$. In this regime, so long as $x_a \gg L^{-(d-d_f)}$, avalanche sizes are truncated by temperature, with $\langle s\rangle \sim x_a^{-\theta}$. 

\section{Conclusions}
We have proposed a scaling description for the stationary flow of thermally excited amorphous solids in the intermittent regime, and supported it by numerical analysis of both a mean field and an elastoplastic model. We show that the physics of the intermittent regime is largely controlled by the thermal activation scale $x_{a} = T^{1/\alpha}f(\dot{\gamma}/T^{1/\alpha})$, which we derive from a simple biased random walker picture. The activation scale $x_a$ serves to collapse all quantities related to avalanches and predicts the flow stress. The activation scale also arises naturally from a modified HL equation, which we argue captures the essence of the intermittent regime, i.e.~the fact that avalanches are no longer cooperative. Contrary to the continuous regime, which obeys a simple Widom scaling ansatz for the rheology, with $\Sigma-\Sigma_c \sim T^{1/\alpha}f(\dot{\gamma}/T^{\psi})$, and $\psi = \beta/\alpha$, the intermittent regime instead exhibits a crossover between driving and continuous flow scaling simply as $\dot{\gamma}\sim T^{1/\alpha}$. This resolves the contradictory observations for the crossover exponent between ref.~\cite{korchinski_dynamic_2022} and refs.~\cite{ferrero_yielding_2021,popovic_scaling_2021}  posed in the introduction. 

It is important to emphasize that all our predictions can be directly verified in molecular dynamics simulations as well as rheological experiments on metallic glasses. Unless strictly athermal quasistatic shear is imposed via energy minimization, these systems are driven at finite rate and temperature. At the level of flow curves, the physics of the intermittent regime manifests as deviations from purely Herschel-Bulkley rheology or logarithmic rate dependence for small shear rates and temperatures (Fig.~\ref{fig:intermittent_rheology}c and d). If the statistics of stress drops are measured for different rates and temperatures, their scaling behavior can be tested against our predictions from fig.~\ref{fig:avalanches}c-e. We hope that the present theoretical development will trigger such work.

Now that we have a handle on the scaling of fluctuations in the intermittent regime, it would be interesting to understand the scaling of temperature on stress-fluctuations in the continuously flowing regime. We would expect that the transition in rheological behaviour at $\dot{\gamma}\sim T^{\beta/\alpha}$, predicting essentially the athermal scaling, with $\xi \sim \dot{\gamma}^{-\beta/\nu}$, while for higher temperatures $\xi$ would be further depressed, though to what extent driving rate and temperature might compete in this regime is not clear. To the best of our knowledge, no one has attempted to directly probe the correlation length in the continuously flowing regime in EPM simulations. Recent work studying dynamical heterogeneity in supercooled liquids has suggested a variant of the four-point correlation function $\chi^4$~\cite{tahaei_scaling_2023}, which might be adapted to serve in the context of driven flow.

Although we (borrowing heavily from \cite{popovic_scaling_2021}) worked out the solution for the thermal HL model with $\alpha=1$, we have not constructed an asymptotic expansion and full boundary layer theory for the model. It would be an interesting project to study the limit of small $T$ perturbatively, as well as working out the scaling close to the critical point of $a = \sigma_c^2/2$. 

Finally, we have used the activation rate: $\lambda(x) = \omega_0 \exp(-x^{\alpha}/T)$, but the scaling of saddle-node bifurcations suggests  $\omega(x) = \omega_0 x^{1/4}$ and $\alpha = 3/2$. This is the starting point used by Chattoraj and Lemaitre \cite{chattoraj_universal_2010}, and considering just the additional thermal activation scale from this arrhenius factor, they arrived at a well-known correction
\begin{equation}
    \Sigma(\dot{\gamma},T) = \Sigma_{\text{athermal}}(\dot{\gamma}) - \left( -T\log(C\dot{\gamma}/T^{5/6}) \right)^{2/3}\,,
\end{equation}
to the Johnson-Samwer law~\cite{johnson_universal_2005}. This was proposed to  account for experimental results on metallic glasses, and works reasonably well in that context. However, this form is not compatible with the simple Widom scaling $\Sigma - \Sigma_c \sim T^{1/\alpha} f(\dot{\gamma} / T^{\beta/\alpha})$ if $\beta=2$. It may be worthwhile to study a variation of the HL model or  EPMs using the activation rule $\lambda(x) = \omega_0 x^{1/4} \exp(-x^{\alpha}/T)$ to understand what effect, if any, this scaling form has on both rheology and avalanching behavior. 

\section{Acknowledgements}
We thank the Natural Sciences and Engineering Research Counci of Canada (NSERC) for financial support. This research was undertaken thanks, in part, to funding from the Canada First Research Excellence Fund, Quantum Materials and Future Technologies Program.

\bibliography{biblio}

\section{Appendices}
\subsection{Solution to the thermal HL model \label{sec:HL_sln}}
The equations defining the thermal HL model are
\begin{equation}
    \partial_t P(\sigma,t) = D\partial_\sigma^2 P - \dot{\gamma}\partial_\sigma P +\Gamma\delta(\sigma) - \nu(\sigma) P 
\label{eq:HL_eqn_app}
\end{equation}
and 
\begin{equation}
    \Gamma = \int P(\sigma)\nu(\sigma) \dup \sigma \,,
\end{equation}
where $\nu(\sigma)$ is the yielding rate of sites with stress $\sigma$, $D = a\Gamma$ is the stress-diffusion due to the activity $\Gamma$, and $\dot{\gamma}$ is the drift due to external loading. $\Gamma$ is the rate at which sites in the system fail, which is why sites are reinjected at $\sigma = 0$ at rate $\Gamma$. This ensures that $\partial_t \int P(\sigma)\dup\sigma = 0$. We consider $\nu = \omega_0\exp(-\text{max}(\sigma_c-\sigma,0)^\alpha/T)$, where $\sigma_c$ is the mechanical stress threshold for sites in the model. For sites with $\sigma>\sigma_c$, the yielding rate is simply $\nu(\sigma>\sigma_c) = \omega_0$. Since $\omega_0$ sets the natural timescale, we adimensionalize in time using: $\tilde{D} = D/\omega_0$, $\tilde{\nu} = \nu / \omega_0$, $\tilde{\Gamma} = \Gamma / \omega_0$, $\tilde{t}= \omega_0 t$, and $\tilde{\dot{\gamma}} = \dot{\gamma} /\omega_0$. The dimensionless HL equations are
\begin{equation}
    \partial_{\tilde{t}} P(\sigma,\tildet) = \tD\partial_\sigma^2 P - \tgdot \partial_\sigma P +\tGamma\delta(\sigma) - \tnu(\sigma) P 
\label{eq:hl_dimensionless}
\end{equation}
and 
\begin{equation}
    \tGamma = \int P(\sigma)\tnu(\sigma) \dup \sigma \,,
\end{equation}

We focus first on the exactly solvable case of  $\alpha = 1$ in the steady-state so that $\partial_t P(\sigma,t) = 0$. Due to the form of $\nu$, as well as the delta-function at $\sigma = 0$, there are discontinuities in the higher derivatives of $P(\sigma)$ at $\sigma = 0$ and $\sigma = \pm \sigma_c$. So, we solve the HL equations piecewise over each interval, and glue them together, requiring continuity everywhere (3 conditions at $\sigma = 0$, $\sigma = \pm\sigma_c$), smoothness at $\sigma = \pm \sigma_c$ (2 conditions), finiteness at $\sigma \rightarrow \pm \infty$ (2 conditions), and normalization to one (1 condition). Since eq.~\eqref{eq:HL_eqn_app} is second order, each reagion gives two solutions. The eight conditions  fully constrain the solution so that there are no remaining degrees of freedom, and there should be a unique solution.  We proceed by solving the HL equations for each region, for some fixed $D$. 

\textbf{Case I: $|\sigma| > \sigma_c$.} This is the simpler case, with the HL equations reducing to simply: 
\begin{equation}
    0 = \tD P'' - \tgdot P' - P 
\end{equation}
This has the solution
\begin{equation}
    P(\sigma)  = A_+ e^{k_+ \sigma} + A_-e^{k_- \sigma}
\end{equation}
where 
\begin{equation}
    k_\pm = \frac{\pm \sqrt{\tgdot^2 + 4 \tD } + \tgdot}{2\tD}\,. 
\end{equation}
The requirement that the solution $P(\sigma\rightarrow \pm \infty) \rightarrow 0$ implies that for $\sigma > \sigma_c$, $A_+ = 0$ and that for $\sigma < -\sigma_c$, $A_- = 0$. Since the constants $A_\pm$ were arbitrary, we can absorb extra factors of $\exp(\pm k_\pm\sigma_c)$, giving the following solutions: 
\begin{equation} P(\sigma<-\sigma_c) = A_+ e^{k_+(\sigma+\sigma_c)}\end{equation} and \begin{equation} P(\sigma>\sigma_c) = A_- e^{k_-(\sigma-\sigma_c)}\,.\end{equation}
Conveniently, the definition of these equations gives that $P(\pm \sigma_c) = A_\mp$. Furthermore, the first derivative at $\pm\sigma_c$ is  simply $P'(\pm \sigma_c) = k_\mp P(\pm\sigma_c) = k_\mp A_\mp$.

\textbf{Case II: $|\sigma| < \sigma_c$.} 
Meanwhile for $\alpha = 1$, the steady-state HL equation becomes for $\sigma \in (-\sigma_c,\sigma_c)\setminus\{0\}$
\begin{equation}
    0 = \tD P'' - \tgdot P' - \exp[-\sigma_c/T]\exp[\ell \sigma/T] P
    \label{eq:HL_region_23}
\end{equation}
where $\ell = \text{sign}(\sigma)$, so that $\ell \sigma = |\sigma|$. Our objective is to transform this equation into the modified Bessel equation: 
\begin{equation}
    0 = z^2 \partial_z^2 \tilde{R}(z) + z \partial_z \tilde{R}(z) - (\kappa^2 + z^2) \tilde{R}(z)
\end{equation}
for which the solutions to $R(z)$ are the modified Bessel functions $K_\kappa(z)$ and $I_\kappa(z)$. Inspired by a change of variables in  ref.~\cite{popovic2021thermalflow}, we use:
\begin{equation}
    \tilde{R}(z) = R(\sigma) = e^{-\tgdot\sigma / 2\tD} P(\sigma)\,. \label{eq:R_defn}
\end{equation}
This leads to 
\begin{equation}
    \partial_\sigma R(\sigma)=R'= e^{-\tgdot \sigma / 2\tD}\left( -\frac{\tgdot}{2\tD} P  +  P' \right)
\end{equation}
and 
\begin{equation}
   \partial_\sigma^2 R(\sigma)  = R'' =   e^{-\tgdot\sigma / 2\tD}\left( \frac{\tgdot^2}{4\tD^2} P  - \frac{\tgdot }{\tD}P' + P''  \right) \,.  \label{eq:R''}
\end{equation}
For $z$, let's try  as an ansatz
\begin{equation}
    z = A e^{L\sigma / T}
\end{equation}
where $A$ and $L$ are, at this point, unknown. Then, 
\begin{equation}
    \partial_z \sigma = \frac{T}{Lz}
\end{equation}
and 
\begin{equation}
    \partial^2_z \sigma = \frac{-T}{Lz^2}\,.
\end{equation}
So,
\begin{equation}
    \partial_z \tilde{R}(z) = R' \partial_z \sigma = \frac{T}{Lz}R'
\end{equation}
and 
\begin{equation}
    \partial_z^2 \tilde{R}(z) =  R' \partial_z^2 \sigma + R'' (\partial_z \sigma)^2 = -\frac{T}{Lz^2}R' + \frac{T^2}{L^2z^2} R''
\end{equation}
We can begin to arrange this in the form of the modified Bessel equation: 
\begin{equation}
    z^2 \partial_z^2 \tilde{R} + z\partial_z \tilde{R} = \frac{T^2}{L^2}R'' = (\kappa^2 + z^2)R(\sigma)
\end{equation}
and inserting eqs.~\eqref{eq:R_defn},\eqref{eq:R''}, we have:
\begin{align}
    &\frac{T^2}{L^2}  e^{-\tgdot\sigma / 2\tD}\left( \frac{\tgdot^2}{4\tD^2} P  - \frac{\tgdot}{\tD}P' + P''  \right)\\ =& (\kappa^2+ A^2 e^{2 L \sigma / T} ) e^{-\tgdot\sigma / 2\tD} P
\end{align}
Now, using eq.~\eqref{eq:HL_region_23}, and cancelling common terms, 
\begin{align}
    &\frac{T^2}{L^2}  \left( \frac{\tgdot^2}{4\tD^2}   + \frac{1}{\tD} e^{-(\sigma_c - \ell \sigma)/T}  \right)P\\=& (\kappa^2+ A^2 e^{2 L \sigma / T} ) e^{-\tgdot\sigma / 2\tD} P
\end{align}
which implies: 
\begin{equation}
L = \ell/2    
\end{equation}
\begin{equation}
    A^2 = \frac{4T^2e^{-\sigma_c / T}}{\tD}
\end{equation}
\begin{equation}
    \kappa^2 = \frac{T^2\tgdot^2}{\tD^2}
\end{equation}
so:
\begin{equation}
    z = \frac{2T}{\sqrt{\tD}}e^{-(\sigma_c- |\sigma|)/2T}
\end{equation}
and the general solution to eq.~\eqref{eq:HL_region_23} is \begin{equation}P(\sigma) = e^{\tgdot \sigma / 2\tD}\left(A_I I_\kappa(z) + A_K K_\kappa(z)\right) \,.\end{equation}
\textbf{Continuity and smoothness at $\pm\sigma_c$.}  We can consider the solution in the vicinity of $\sigma = \pm \sigma_c$, and write $P(\pm \sigma>0) = P_{\mp}(\sigma)$ where: 
\begin{equation}
    P_{\mp}(\sigma) = e^{-\frac{\tgdot(\sigma_c\mp \sigma)}{2\tD}}\left(A_{I,\mp} I_\kappa(z_\mp)+A_{K,\mp} K_\kappa(z_\mp) \right)
\end{equation}
and
\begin{equation}
    z_{\mp}(\sigma) = \frac{2T}{\sqrt{\tD}}e^{-(\sigma_c\mp \sigma)/2T}\,.
\end{equation}
Now, at $\sigma = \pm\sigma_c$, we have that $z_\mp(\pm\sigma_c) = \frac{2T}{\sqrt{\tD}}$ and that $z'_{\mp}(\pm\sigma_c) = \frac{\pm1}{\sqrt{\tD}}$. Using the results from case I: $P(\pm \sigma_c) = A_\mp$ and $P'(\pm \sigma_c) = k_\mp P(\pm\sigma_c) = k_\mp A_\mp$, with the results from case II, we can concisely express the continuity and smoothness conditions as a matrix:
\begin{equation}
\begin{bmatrix}
    1 \\ \sqrt{\tgdot^2 + 4\tD}
\end{bmatrix}
= 
\begin{bmatrix}
   I_\kappa\left(\frac{2T}{\sqrt{\tD}}\right) &K_\kappa\left(\frac{2T}{\sqrt{\tD}}\right) \\
    \frac{-1}{\sqrt{D}}I'_\kappa\left(\frac{2T}{\sqrt{\tD}}\right) &\frac{-1}{\sqrt{D}}K'_\kappa\left(\frac{2T}{\sqrt{\tD}}\right) \\
\end{bmatrix}
\begin{bmatrix}
    \dfrac{A_{I,\mp}}{A_{\mp}}\\ \dfrac{A_{K,\mp}}{A_{\mp}} 
\end{bmatrix}\label{eq:matrix_expression_AIAK_over_A}
\end{equation}
These equations can be inverted to obtain the ratios $A_{I,\pm} / A_\pm$ and $A_{K,\pm} / A_\pm$.

\textbf{Continuity at $\sigma = 0$. } We can use continuity at $\sigma = 0$ to relate $A_-$ to $A_+$. At $z_0 = z(0) = \frac{2T}{\sqrt{\tD}}e^{-\sigma_c/2T}$. 
\begin{equation}
\frac{P_\pm(0)}{A_\pm} = e^{-\frac{\tgdot\sigma_c}{2\tD}}\left(\dfrac{A_{I,\pm}}{A_{\pm}} I_\kappa(z_0) +\dfrac{A_{K,\pm}}{A_{\pm}} K_\kappa(z_0)\right) 
\end{equation}
Now, since $A_{I,-}/A_- = A_{I,+}/A_+$  and $A_{K,-}/A_- = A_{K,+}/A_+$, we can rearrange to obtain:
\begin{equation}
A_+/A_-= \exp(-\tgdot \sigma_c /\tD) \label{eq:AplusoverAminus}
\end{equation}

So, with eq.~\ref{eq:AplusoverAminus},\ref{eq:matrix_expression_AIAK_over_A} in hand, a full solution of the form $P(\sigma)/A_-$ can be written, using the coefficients $A_{I/K,\pm}/A_-$, and $A_+/A_-$. The normalized solution can be worked out by applying the final condition: $\int P(\sigma)\dup \sigma = 1$, which fixes the value for $A_-$.

\begin{figure}
    \centering
    \includegraphics{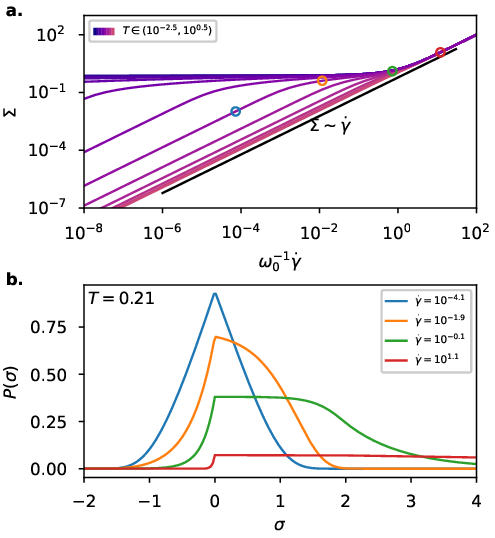}
    \caption{Rheology of the  $\alpha = 1$ thermal HL model. }
    \label{fig:suppl_rheology}
\end{figure}

\textbf{Asymptotic scaling of $\langle\sigma\rangle$.}  
In the limit of low strain-rates and intermediate temperatures, the bulk of $P(\sigma)$ is in $|\sigma|<\sigma_c$. In this case, for $\sigma < 0$, the solution is quite close to $p(\sigma) \sim e^{\tgdot\sigma/\tD}$. For $\sigma > 0$, the solution is dominated by $P(\sigma) \sim K_\kappa(z) e^{\tgdot\sigma / 2\tD}$. Most of the scaling can be captured as: $K_\kappa(z) \approx z^{-\kappa} \exp\left[-z\right]$, and of that scaling, most of it can be captured as $P(\sigma)\sim \exp(-z)$, so 
\begin{equation}
P(\sigma>0) \approx \exp\left[ -\frac{2T}{\sqrt{\tD}}e^{-(\sigma_c-\sigma)/2T}\right]\end{equation} 
which is approximately constant for $\sigma < \sigma^*$ and very quickly goes to zero at $\sigma = \sigma^*$, for $\sigma^*$ satisfying 
\begin{equation}
    1 = \frac{2T}{\sqrt{\tD}}e^{-(\sigma_c - \sigma^*)/2T}\end{equation}
    \begin{equation} \implies\sigma^* =  \sigma_c- 2T\log\left(\frac{2T}{\sqrt{\tD}}\right)\,.
\end{equation}
Now, to compute the flow stress, we have   (using $
tD = \frac{1}{2}a \tgdot$)
\begin{equation}\langle \Sigma \rangle = Z \left[\int_{-\infty}^0\sigma e^{2\sigma / a}  \dup \sigma  + \int_0^{\sigma^*} \sigma \dup \sigma \right] = Z\left[ \frac{1}{2} (\sigma^*)^2 - \frac{a^2}{4} \right] \,, \end{equation} 
where $Z^{-1} =  \left[\int_{-\infty}^0 e^{2\sigma / a}  \dup \sigma  + \int_0^{\sigma^*} 1\dup \sigma  \right] = \frac{a}{2} + \sigma^*$ is the overall normalization.  For sufficiently small $a$, this gives $\langle \Sigma\rangle = \frac{1}{2}\sigma^*$, and therefore
\begin{equation}
    \langle \Sigma \rangle \approx \frac{1}{2}\left(\sigma_c - 2 T \log\left[\frac{2T}{\sqrt{\frac{1}{2}a \tgdot }}\right] \right)
\end{equation}
which should be valid in the limit of small(ish) strain rates, when we don't have appreciable $P(\sigma > \sigma^*)$. 

However, for sufficiently small $\dot{\gamma}$, this cannot apply, because eventually $\sigma_c <  2 T \log\left[\frac{2T}{\sqrt{\frac{1}{2}a \tgdot}}\right]$, and the overall stress would become negative. We find that at a very low temperature-dependent strain-rate, the stress-strain relation actually becomes linear (cf. fig.~\ref{fig:suppl_rheology}a), at which time the site-stress distribution $P(\sigma)$ becomes very nearly symmetric around zero. We treat treat this case in the next section.

\textbf{High-temperature $\beta = 1$ limit.} For sufficiently hot, or slowly driven, systems, the flow stress becomes small, and the distribution of stresses becomes very nearly symmetric (cf. bottom panel of Fig.~\ref{fig:suppl_rheology}). 

For the high-temperature case, we can approximate the HL equation by: 
\begin{equation}
    \tD P'' - \tgdot P' - P = 0
\end{equation}
which amounts to assuming that the activation rate is independent of temperature. For such a case, $\tD = a\int P = a$. Then, the solution everywhere is just given by $P(\pm \sigma >0) = A e^{k_{\mp}\sigma}$. The normalization factor $A$ is then simply $A^{-1} =  -k_-^{-1}  + k_+^{-1}$. To obtain the stress, direct integration yields 
$\langle \Sigma \rangle = A\left(-1/k_+^2 + 1/k_-^2\right) = \tgdot / 2$, yielding $\beta = 1$. This is observed in mettalic glasses operated at low stresses near or below their glass transition~\cite{homer_mesoscale_2009}.

\textbf{$\beta = 2$ to $\beta=1$ crossover.} The athermal HL model, for $a < \sigma_c^2/2$~\cite{olivier_glass_2011,agoritsas_non-trivial_2017}, is a yield stress fluid, with a low strain-rate rheology: 
\begin{equation}
    \Sigma = \Sigma_c + c_1 \dot{\gamma}^{1/2} + c_2\dot{\gamma} \,.
\end{equation}
The athermal model is always asymptotically $\beta = 2$ for small strain rates, since $c_1 \sqrt{\dot{\gamma}} > c_2 \dot\gamma$ for strain rates below $\dot{\gamma} < (c_1/c_2)^2$. At higher strain rates, the stress-strain relation is linear, as is also the case in the thermal model at high strain rates (cf.~fig.~\ref{fig:suppl_rheology}). However, the prefactors $c_1$ and $c_2$ depend on the cooperativity factor $a$.  This is why for sufficiently low $a$, the thermal model can exhibit a purely $\beta = 1$ rheology, above the logarithmic scale $T^{\beta/\alpha}$. To be explicit, if $(c_2(a)/c_1(a))\sim a^p \sim \dot{\gamma} < T^{\beta/\alpha}$, we would expect to only see the $\beta = 1$ behaviour whenever $a < T^{\beta/(p\alpha)}$. This naturally connects to the non-interacting thermal HL model, where $a=0$, treated in the next section. 

\subsection{Solution to the non-interacting thermal HL model \label{sec:niHL_sln}}
The non-interacting HL model is defined by the Fokker-Planck equation:
\begin{equation}
    \partial_{\tildet} P(\sigma,t) = -\tgdot\partial_\sigma P - \tnu(\sigma)P(\sigma) 
\end{equation}
As before, we'll solve this in equilibrium, for the regions $\sigma \in (0,\sigma_c)$ and $\sigma > \sigma_c$ separately, and then glue solutions in the different domains together to form a complete solution. 

\textbf{Case I: $\sigma \in (0,\sigma_c)$}in equilibrium this is
\begin{equation}
    \tgdot P' = \exp[-(\sigma_c-\sigma)^\alpha/T]P
\end{equation}
This has the solution
\begin{equation}
    \log(P(\sigma)/P(\sigma = 0)) = -\int_0^\sigma \frac{1}{\tgdot} \exp[-(\sigma_c-\sigma)^\alpha/T]\dup \sigma  
\end{equation}
making the substitution $u = (\sigma_c-\sigma)^\alpha/T$ yields
\begin{equation}
    \log\left(\frac{P(\sigma)}{P_0}\right) = \int\displaylimits_{\sigma_c^\alpha/T}^{(\sigma_c-\sigma)^\alpha/T}\frac{1}{\alpha\tgdot} T^{1/\alpha}  u^{1/\alpha-1}\exp[-u]\dup u
\end{equation}
\begin{equation}
    \log\left(\frac{P(\sigma)}{P_0}\right) = \frac{-T^{1/\alpha}}{\alpha \tgdot }\Gamma\left(\frac{1}{\alpha},\frac{(\sigma_c-\sigma)^\alpha}{T},\frac{\sigma_c^\alpha}{T},\right)
\end{equation}
where $\Gamma(a,z_0,z_1)$ represents the generalized incomplete gamma function defined $\Gamma(a,z_0,z_1) = \int_{z_0}^{z_1} z^{a-1}e^{-z}\dup z$. So
\begin{equation}
    P(\sigma) = P_0\exp\left[\frac{-T^{1/\alpha}}{\alpha \tgdot }\Gamma\left(\frac{1}{\alpha},\frac{(\sigma_c-\sigma)^\alpha}{T},\frac{\sigma_c^\alpha}{T},\right)\right]\label{eq:ni_soln_0}
\end{equation}

\textbf{Case II: $\sigma > \sigma_c$.} Here the HL equation is simply $\tgdot P' =  -P$, with an exponential decay as the solution, with $P(\sigma) = \exp[-\sigma/\tgdot]$. 

\textbf{Scaling form for $\langle\Sigma\rangle$.}
When $T$ is large compared to $\tgdot$, $P(\sigma_c)\rightarrow 0$, and when computing $\langle \Sigma\rangle$ we  need only concern ourselves with the part of the solution given by eq.~\eqref{eq:ni_soln_0}. This solution is essentially constant, until a sharp transition at $\sigma^* = \sigma_c-x_a$. The slope in $P(\sigma)$ is maximized when:
\begin{equation}
    0 = \partial^2_\sigma P = \partial_\sigma \frac{-\tnu}{\tgdot}P = (-\frac{1}{\tgdot} \partial_\sigma \tnu) P + \frac{\tnu^2}{\tgdot^2}P 
\end{equation}
and since $\partial_\sigma \tnu = \alpha (\sigma_c-\sigma)^{\alpha-1}/T \tnu$, and since we are solving for the point of maximum slope, $\sigma^* = \sigma_c - x_a$:  
\begin{equation}
\implies \alpha x_a^{\alpha-1}\frac{\tgdot}{T} = e^{-x_a^\alpha/T} \label{eq:x_a_eqn_ni_hl}\,.
\end{equation}
This is precisely the same equation we obtained when considering the activation threshold for a single random walker, and will give the same $\langle x_a\rangle$ given by eq.~\eqref{eq:xaavg}. 
Simple algebraic rearrangement of eq.~\eqref{eq:x_a_eqn_ni_hl} yields
\begin{equation}
    -\frac{\alpha}{1-\alpha} \left( \alpha \tgdot/T\right)^{\alpha/(1-\alpha)}\frac{1}{T} = -\frac{\alpha}{1-\alpha}\frac{x_a^\alpha}{T}\exp\left[-\frac{\alpha}{1-\alpha}\frac{x_a^\alpha}{T}\right]
\end{equation}
Now, the Lambert-W function $\lambW(z)$ satisfies $z =\lambW(z)e^{\lambW(z)}$, so if we take $z = \frac{\alpha}{1-\alpha}\frac{x_a^\alpha}{T}$ and $\lambW(z) = -\frac{\alpha}{1-\alpha}\frac{x_a^\alpha}{T}$, we obtain:
\begin{equation}
    \lambW\left(\frac{\alpha}{\alpha-1} \left(\frac{T^{1/\alpha}}{\alpha\tgdot}\right)^{\alpha/(\alpha-1)}  \right) = \frac{\alpha}{\alpha-1}\frac{x_a^\alpha}{T}
\end{equation}
which yields eq.~\eqref{eq:xaavg}.

\subsection{Functional form for $P(x_{a})$ \label{sec:pxa_discussion}}
Here we work out the functional form for $P(x_a)$ for a walker beginning at $x_0 = \infty$. 
\begin{equation}
    P(x_a) = \frac{1}{\dot{\gamma}}\lambda(x_a)\exp\left[-\frac{1}{\dot{\gamma}}\int^{\infty}_{x_a} \lambda(x') dx'\right]\,.
\end{equation}
For $\alpha = 1$, 
\begin{equation}
    P(x_a) = \frac{1}{\omega_0^{-T}\dot\gamma}\exp\left(\frac{-1}{\omega_0^{-1}\dot\gamma} e^{-x_a/T} \right) \exp(-x_a/T) 
\end{equation}
and for $\alpha = 2$, this is
\begin{equation}
     P(x_a) =\frac{1}{\omega_0^{-1}\dot\gamma}\exp\left(\frac{-\sqrt{\pi T/2}}{\omega_0^{-1}\dot\gamma}\text{erf}(-x_a/T^{1/\alpha}) \right) \exp(-x_a^2/T) 
\end{equation}
These probability distributions are not normalized (over the domain $x_a (0,\infty)$, once we go to high enough driving rate.  That is because a finite fraction of sites will escape to $x=0$, where they are mechanically excited, as opposed to thermally excited. The normalization constant $f = \int P(x_a) dx_a$, indicates the fraction of sites that are still thermally excited before reaching $x=0$.

\subsection{Functional form of  $p(x)$ \label{sec:px_discussion}}
In the intermittent regime with finite-temperatures, $p(x)$ becomes gapped below the activation scale $x_a(T,\dot{\gamma})$. We describe in the text the scaling for $\xmavg$ when $x_a > L^{-(d-d_f)}$, i.e. when the gap interrupts the pseudogap, as well as when $x_a < L^{-(d-d_f)}$, when it interrupts the slowly slowly varying region.

\textbf{Gapping the pseudogap: $x_a > L^{-(d-d_f)}$.} Here we will consider the effect of different functional forms for $p(x)$, and derive the scaling forms presented in the main text. For the case that $x_a > L^{-(d-d_f)}$, we expect a pseudogap scaling $p(x)\sim x^{\theta}$ for $x\gg x_a$ and $x\ll 1$, with $P(x < x_a)=0$. However, there are many scaling forms that satisfy this relation. The simplest, and the one we use to derive the results in the main text, is simply $p(x>x_a) = c x^\theta $ for some normalization $c$, and $P(x<x_a) = 0$. This distribution has a discontinuity as $x=x_a$. Another option, one that was considered in \cite{popovic2021thermalflow}, is $p(x)\sim (x-x_a)^\theta$. This second options has the appealing feature that it effectively shifts $x=0$ to $x=x_a$.

Although these two forms for $p(x)$ both satisfy the obvious requirements on $P(x)$, the difference in scaling close to $x=x_a$ actually has consequences for the scaling of $\xmavg$, and more importantly, $\langle s\rangle L^{-d} = \xmavg-\xaavg$. For $p(x)\sim (x-x_a)^\theta$, using $L^{-d} \sim \int_0^\xmavg p(x) \dup x$, one finds $\xmavg-\xaavg = k L^{-d/(1+\theta)} $. This would imply that avalanche sizes are strictly controlled by system size, and that $x_a$ plays no role, contrary to our findings in fig.~\ref{fig:avalanches}c. 

If we instead use $p(x) = c x^\theta$, then $L^{-d} \sim \xmavg^{\theta+1} - x_a^{\theta+1}$, so $\xmavg = \left(kL^{-d}+x_a^{1+\theta} \right)^{1/(1+\theta)}$. If $L^{-d} \ll x_a^{1+\theta}$, then $\xmavg-x_a\sim L^{-d}x_a^\theta$ to lowest order, giving the result of eq.~\ref{eq:s_avg_T_scaling_regime}.  

\textbf{Gapping the plateau: $x_a < L^{-(d-d_f)}$.} For $x<L^{-(d-d_f)}$, the stability distribution deviates from a power-law. In brief, this is because after an avalanche of size $S$, the stability distribution acquires a plateau below a stress scale $sL^{-d}$~\cite{korchinski_signatures_2021}. When averaging the stability distribution sampled after many avalanches, this results in a plateau below $L{-d}$ (from avalanches corresponding to $s=1$). The deflection from power-laws begins at $L^{-(d-d_f)}$, since  that is the scale associated with the largest avalanches. Between the scales $L^{-d}$ and $L^{-(d-d_f)}$, the stability distribution follows a shallow, size dependent, powerlaw $\tilde{\theta}(L)$. For progressively larger systems, the largest avalanches play an outsized role, and  $\tilde{\theta}(L)$ approaches zero. So, we approximate $p(x)$ as a plateau for $x<L^{-(d-d_f)}$.

If both $\xmavg$ and $\xaavg$ are in the plateau, then $L^{-d} \sim \int_{x_a}^\xmavg p(x)\dup x \sim L^{-p}(\xmavg-\xaavg)$, which implies a size-dependent avalanche cut-off, with $\langle s\rangle \sim L^{d-p}$. This is the same avalanche size expected for the athermal system~\cite{korchinski_signatures_2021}, implying that when both $\xaavg$ and $\xmavg$ are in the plateau, there is essentially no effect of temperature on avalanche size.

In principle, it could be the case that $\xaavg < x_c= L^{-(d-d_f)} < \xmavg$. This would suggest additional corrections very close to the crossover, with $\xmavg$ satisfying $c_2L{-d} = c_2 (x_c-x_a)L^{-p} + \xmavg^{\theta+1}-x_c^{(1+\theta)}$. However, these corrections would only apply when $x_a \approx x_c = L^{-(d-d_f)}$, and appear to be small enough we cannot see them in our EPM data.

\end{document}